**UK Space Frontiers 2035 Science White Paper**

# The SPace-based InterFerometer Feasibility (SPIFF) Project: Enabling Future High-Resolution Astronomy Across the EM Spectrum

Thematic Area: Astronomy and Astrophysics

**Lead author:**

Name: Berke Vow Ricketti

Institution: Disruptive Space Technology Centre, RAL Space, STFC, UKRI

Email: berke.ricketti@stfc.ac.uk

**Co-authors:** (names and institutions)

Victoria Yankelevich (RAL Space, STFC), Chris Benson (Cardiff U.), Renske Smit (LJMU), Sebastian Kamann (LJMU), Ettore Pedretti (RAL Space, STFC), Sebastian Marino (U. Exeter), Gerard van Belle (Ad Astra Space LLC, USA), Stephen Eales (Cardiff U.), Chris Bee (Oxford Space Systems), Mark Wyatt (U. Cambridge), Matthew Smith (Cardiff U.), Tim D. Pearce (U. Warwick), Emily Williams (RAL Space, STFC), Rebecca Harwin (RAL Space, STFC), David Pearson (RAL Space, STFC), Andy Vick (RAL Space, STFC), Giorgio Savini (UCL), Taro Matsuo (U. Osaka, Japan), Hiroshi Matsuo (NAOJ, Japan), Locke Spencer (U. Lethbridge, Canada), and David T. Leisawitz (NASA Goddard, USA)

**Abstract:** A plethora of astronomical science cases can only be achieved with high angular resolution observations, and we can expect the number of these to grow as astronomers are constrained by the size limitations of monolithic single-aperture space telescopes, making space-based interferometry inevitable. However, due to complex engineering challenges, the enabling technologies do not have flight heritage at the system level, and the concept of space-based interferometry is still immature in the eyes of the broader astronomical community, meaning ***no direct-detection synthetic-aperture space-based interferometer has yet flown*** and an opportunity exists for the UK to take a world leading role. Here we propose the SPace-based InterFerometry Feasibility (SPIFF) Project as a program to address both issues simultaneously by: 1) completing a thorough survey of the science cases across the EM spectrum that would directly benefit from, or be impossible without, space-based interferometry; 2) down selecting the key requirements via a Science Traceability Matrix mapping exercise to these science cases, allowing us to identify technology hurdles; 3) host a workshop for the UK astronomical community to consolidate these findings and raise confidence in concept maturity; 4) build a technology demonstration mission to raise TRL and achieve flight heritage of critical technologies, alleviating any lingering scepticism around the concept of space-based interferometry. Such a program positions the UK as the ***partner of choice*** for any future NASA or ESA space-based interferometry mission, allowing the UK to lead groundbreaking scientific discoveries, while also directly benefiting the UK industrial base by advancing domestic exportable technologies and building direct synergy with other UK space priorities like SDA, ISAM, and PNT. Indeed, the UK is uniquely positioned to lead in space-based interferometry, possessing a rare trifecta of domestic strengths: world-class expertise in ground-based interferometry and space-based instrumentation; commercial entities developing mission-critical technologies; and scientists whose research spans the full range of science cases that would benefit directly from space-based interferometry. In many respects, this represents not merely a Decadal opportunity, but a ***Generational*** one.

## 1. Scientific Motivation and Objectives

*"Are we alone?" "How did we get here?" "How does the universe work?"*
~ NASA Astrophysics Roadmap, "Enduring Quests, Daring Visions", 2014[1]

Groundbreaking discoveries require groundbreaking observations. Answering the Enduring Questions of Humanity requires astronomical observations beyond the capabilities of any past, present, or planned facility. Over the generations, the Great Observatories of humanity, like the Hubble Space Telescope (*HST*) and the James Webb Space Telescope (*JWST*), have changed our understanding of the universe, re-writing textbooks, and inspiring the next generation of scientists with stunning images of the cosmos. Indeed, the next seismic shift in humanity's scientific understanding could come from the Search for Extraterrestrial Life, Mapping ExoEarths and black holes, and probing deeper into the universe to understand the formation of planetary systems and evolution of galaxies. However, as we evaluate these science cases more closely, it quickly becomes apparent that they all ***require*** high angular resolution observations beyond those capable of single-aperture facilities (Figure 1).

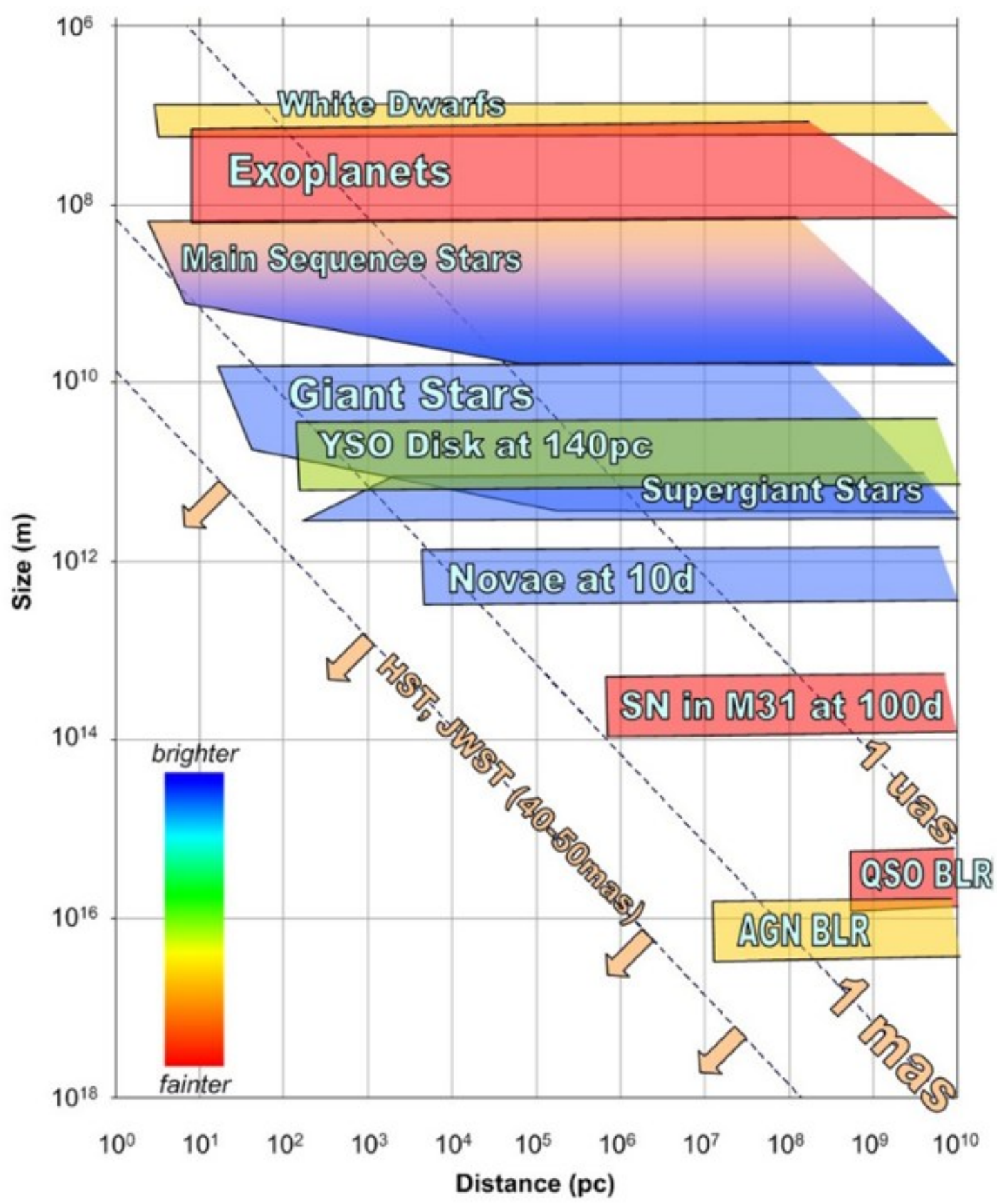

*Figure 1. Astrophysical target size vs. distance. Many astronomical objects of interest are hidden beyond our current observing capabilities. While ground-based interferometry has started to push this boundary, the next obvious step would be space-based interferometry. Credit: Rinehart et al 2019* [32]

**The Search for Extraterrestrial Life, Mapping ExoEarths, and Cataloguing ExoPlanets**: The search for life beyond the Solar System demands capabilities beyond single-aperture telescopes. While missions such as ESA's *Ariel* and NASA's planned Habitable Worlds Observatory (*HWO*) will deliver transformative spectra of exoplanet atmospheres, their monolithic architectures limit spatial resolution, reducing even nearby terrestrial planets to point sources. The natural successor is a space interferometer, building on concepts such as *TPF-I*[2,3], *Darwin*[4,5], and *LIFE*[6,7]. Interferometry in the mid-infrared (mid-IR) would enable direct characterisation of rocky Earth-like planets and the detection of key atmospheric biosignatures. The ultimate goal — to image an ExoEarth well enough to distinguish oceans and continents — requires baselines of hundreds of kilometres, achievable only through formation-flying interferometry. While ambitious, and requiring pushing technological boundaries, a national roadmap is an essential starting point for the UK to contribute to, and potentially lead, this long-term vision.

**Black Holes:** Black holes represent our Universe at its most extreme. In the public consciousness, they lie at the intersection of science fiction and science fact. Indeed, the silhouette of M87 from the Event Horizon Telescope catapulted astrophysics to the forefront of the global public consciousness. However, the final stages in the evolution of massive stars from which black holes form are still poorly constrained. Owing to uncertainties surrounding binary interactions or stellar winds, it is still uncertain which stars collapse into black holes and what the black-hole mass function resulting from massive stellar evolution looks like[8]. Such uncertainties severely limit our abilities to interpret the growing number of gravitational wave detections from merging black holes. One exciting possibility to make progress is via the study of BHs in orbits around luminous stars, such as those discovered in star clusters[9] or by Gaia[10]. However, secure confirmation of black hole candidates around luminous stars requires careful studies of the binary systems to exclude so-called "BH imposters"[11]. Interferometry provides a unique possibility to detect imposters[12]. Indeed, direct detection and impacting imaging of black hole structure requires spatial resolution on scales that can only be practically achieved through interferometry. The Black Hole Mapper, outlined in the Astrophysics Roadmap[1], would provide high angular resolution at x-ray wavelengths, allowing us to explore the innermost regions of accretion discs and gain new insights into the extreme physical regime surrounding black holes. Indeed, probing the physics of black holes with interferometry, as well as the advancement of x-ray interferometry, were amongst the final recommendations of ESA's Voyage 2050 report[13].

**How planetary systems form and evolve:** Beyond finding and studying exoplanets, the methods by which they form and evolve are not yet fully understood. The study of debris discs and protoplanetary discs has greatly benefited from the high-resolution observations and synergetic combination of *ALMA* and *JWST*[14]. However, a gap remains at far-IR wavelengths where no past, present, or planned facility can deliver sub-arcsecond angular resolution at these wavelengths, in which the early stages of protostar collapse and protoplanetary[15] and debris discs[16] are at their brightest. This spectral regime allows observations of water and water-ice features[17], that are essential for understanding the clumping of dust grains, the flow of volatile elements, and the composition of early exo-planet atmospheres[18–20] (particularly as potential biospheres). Large single-aperture far-IR telescopes have been proposed (e.g., *Origins*[21]*, AtLAST*[22]), but would not provide the needed angular resolution. Resolved imaging of debris disc structures around the nearest stars at au scales will be able to trace planetary system architectures in otherwise inaccessible regimes, as well as the history of dynamical interactions between the components[14,23].

**Early Dust Formation**: *JWST* and *ALMA* have started to open a window into the first dust formation, just a few hundred million years of cosmic time. Yet some of the most distant dusty galaxies have been shown to have high dust temperatures peaking near 200µm[24], challenging our understanding of the dust properties of these early galaxies. However, these wavelengths are inaccessible to *ALMA* and the sources are too faint for *Herschel*'s or *PRIMA*'s sensitivity. Furthermore, surprising observations of the 2175Å bump out to redshift z~7, just 800 million years after the Big Bang[25,26], challenge the models of early dust formation. These Polycyclic aromatic hydrocarbons (PAH) features should appear at 25–90 µm, a regime beyond *JWST*'s reach. However, previous far-IR interferometer concepts such as *SPIRIT/SPICE*[18,27,28], *FISICA*[29–31], and *FIRI*[19] offer both the sensitivity of multiple cryogenic apertures and the angular resolution needed to beat the confusion noise limit, enabling direct detection of hot dust and PAH emission from these redshifted galaxies[32] (Figure 2).

**Tracing the earliest stages of star formation:** The far-IR allows access to a multitude of molecular rotational transitions and atomic fine structure lines that, together, provide key tracers of the warm molecular interstellar medium (ISM). Through these, cooling processes, ionisation regions, and shocks that are important drivers in star formation (and ultimately planetary systems) can be traced[33]. The *Herschel* space observatory revolutionised our understanding of star formation in our galaxy through its large aperture. However, more recent studies from the RIOJA project have shown that *JWST* and *ALMA* alone cannot identify the physical condition of ISM[34–36], highlighting the need for high angular resolution far-IR observations that can be provided by a space-based interferometer.

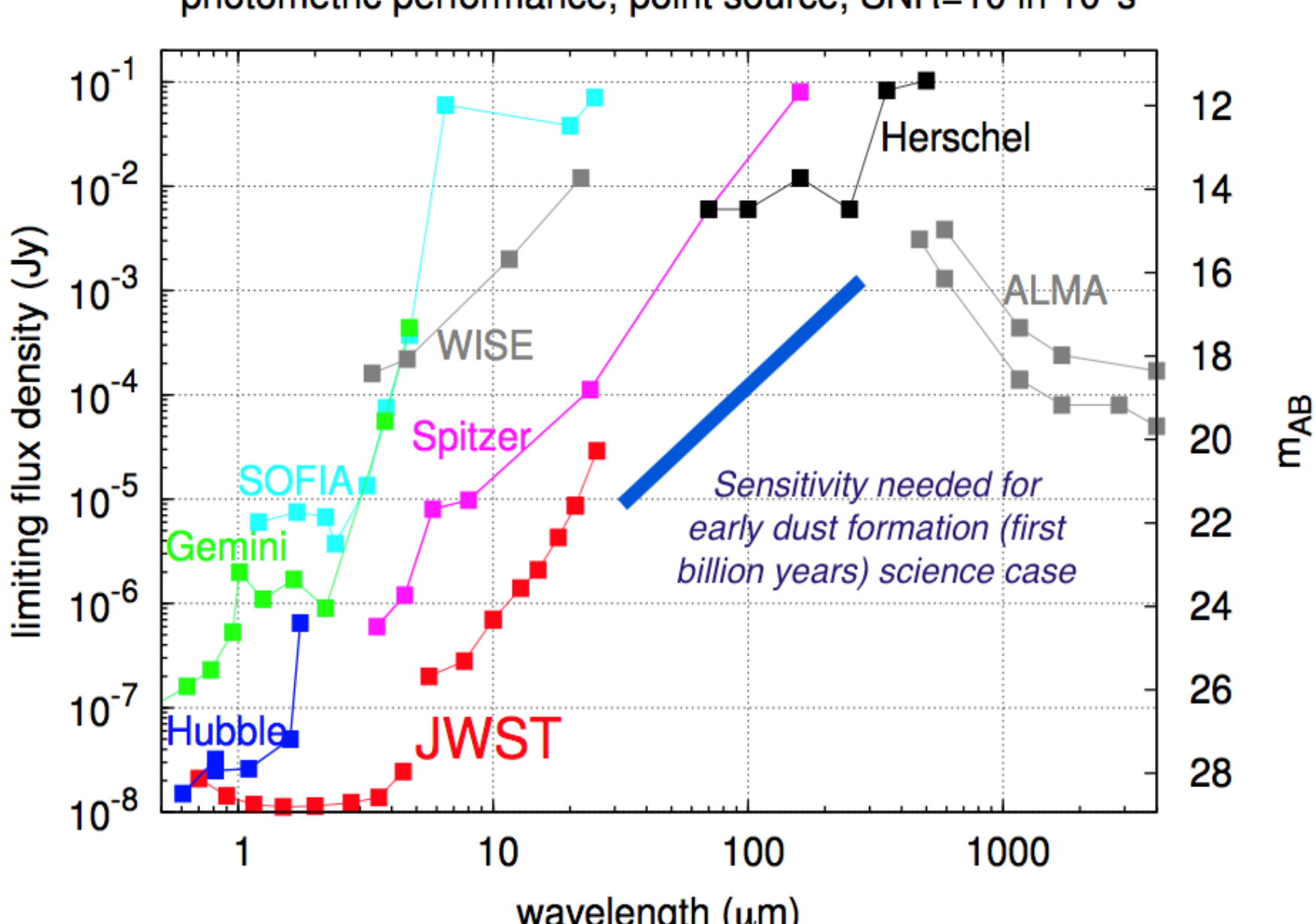

*Figure 2. Sensitivity required for early dust formation science case. A gap in spectral cover exists that no previous, present, or planned facility can fill, but far-IR interferometry mission could address this gap directly. Credit: Adapted from NASA/STScI via George Rieke*

**Galaxy evolution:** Mapping the distribution of star formation activity in active galaxies or observing high-redshift galaxies whose wavelengths have pushed above the observing capabilities of *JWST* would benefit directly from a far-IR space-based interferometer. Indeed, if we manage to match the resolution of far galaxies with the Milky Way, we can investigate star-formation in a far greater range of environments, such as low-metallicity or different morphologies. Moreover, roughly half the observed flux from galaxies is observed in the far-IR[18]. The early results with the *JWST*[37] show the need for a space-based interferometer. The *JWST* surveys show that galaxies at z~10 are only a few hundred pc in size and barely resolvable with the *JWST*. They also suggest that the formation of super-massive black holes predated the formation of galaxies[38], and connection between the two will only be possible to probe with a space-based interferometer.

The science cases outlined above demand angular resolutions far beyond any planned observatory. Monolithic telescopes have reached practical limits—*JWST* already required a segmented deployable 6.5m mirror—and achieving space-based apertures on the scale of 10s of metres to kilometres is only possible through aperture synthesis interferometry. Atmospheric turbulence and opacity further preclude ground-based solutions, especially for far-IR science cases, making a space-based approach essential. Space-based interferometry has been studied at the concept level for decades, but no mission has yet come to fruition. Given this gap in the field, we see an opportunity for the UK to take a leading role in space-based interferometry, with the starting point being the **SPace-based InterFerometry Feasibility (SPIFF) Project**.

In Section 2, we detail the strategic context and gap in global leadership that SPIFF will capitalize on, highlighting direct benefits to the UK. Then, we explain the proposed SPIFF idea in more detail, emphasizing how SPIFF is unique compared to previous efforts in this area. Finally, we explore why the UK is uniquely suited to take a leading role by showcasing existing UK expertise and industrial capability, as well as strategic international partnerships for this effort.

## 2. Strategic Context

*"Several of the notional missions listed in our roadmap rely on interferometry to answer key science questions, from radio to X-rays. All notional missions in the Visionary Era are interferometers, and technology maturation of interferometric techniques is this highly relevant to realizing the science vision."*

~ NASA Astrophysics Roadmap, "Enduring Quests, Daring Visions", 2014[1]

Space-based interferometers have been studied at the concept level for over two decades, from structurally connected space-craft with modest Probe/M-class budgets to a fleet of formation flying collectors on the scale of a Great Observatory/L-class mission, and many more besides[2,5,6,18,19,39–50]. Indeed, the list of concepts cited here is by no means exhaustive and a complete accounting of mission concepts and community workshops is beyond the scope of this paper. The sheer volume of science cases which could benefit from high angular resolution (a small fraction of which were detailed in Section 1) makes space-based interferometry inevitable[51]. Indeed, the astronomical community's enthusiasm for higher angular resolution is evidenced by both the NASA Astrophysics Roadmap[1] and the ESA Voyage 2050[13] emphasizing the importance of space-based interferometry for future observatories across the EM spectrum. Roadmaps towards this future of space-based interferometry, whether focusing on formation flying specifically or more general considerations, have also been published in the past[44,45,52]. However, ***no direct-detection synthetic-aperture space-based interferometer has ever flown***.

Today, while the dream of space-based interferometry is still alive, only a few mission concepts are currently under active study, including *STARI*[53], *SERIOS*[54], *SILVIA*[55], and *LIFE*[6]. *STARI*, an in-orbit formation-flying test, and *SEIRIOS*, the first in-orbit fringe-demonstration mission, have now been approved and are scheduled for launch in 2028 and 2031, respectively. While balloon-borne missions like *BETTII*[56,57] and *JUSTIINE*[58] have also been considered in the past, the path towards a future observatory-class space-based interferometric facility is not yet clear.

***Now*** is the time for the UK to take the lead and pave the path towards an observatory-class mission, and doing so will have direct societal and industrial benefits to the UK, as well as broader 'Big Picture' strategic and economic benefits. UK participation in international science projects, especially those with flagship visibility (e.g. the *HST* Deep Field image or the M87 silhouette), drives public engagement, builds support for STEM programs, and inspires the next generation of scientists and engineers. Furthermore, UK participation is an opportunity to elevate the global profile of UK-based exportable technologies from academic spin-out small businesses, as well as further advancing technologies that have already been developed by UK-based industrial partners. Whether building a technology demonstrator mission worth ~£10 million, resulting in 10s of FTEs for a few years, or a Great Observatory-class mission (e.g. the $3-5 billion budget as recommended for the next far-IR Flagship mission[59]) yielding potentially 100s of FTEs for a decade or more, the direct economic impacts are potentially enormous as high-value jobs are brought to the UK. The scientific output would likewise lead to high-impact research publications, potentially earning international awards and initiating a positive feedback loop that raises the UK's global scientific profile, increases public support for scientific funding and education, and driving further technological advancement.

This promising future is ***only*** achievable through a coherent, well-crafted roadmap for space-based interferometry—beginning with a systematic mapping of science cases to technological capabilities, a survey of the opportunity landscape, the development of Science Traceability Matrices (STMs)[60,61] and technology trade studies, and building a technology demonstration mission to showcase concept maturity, all anchored in UK leadership. The SPIFF project represents the pivotal first step in this journey, laying the foundations for a future observatory-class space-based interferometer mission.

## 3. Proposed approach

*"While the scientific drivers for space-based interferometry are compelling, to attain this capability we need to have a path from the present to this future."*

~Rinehart et. al, "A Long-Term Vision for Space-Based Interferometry", 2019[32]

For decades, space-based interferometry has faced an uphill battle in both technical challenges and broader community perception. Low TRLs for mission-critical technologies, lack of flight heritage at the system level, and a general perception of concept immaturity has left space-based interferometry to languish for years while single-aperture concepts continue to advance across the EM spectrum. For example, beam combination and fringe tracking on ground-based optical interferometers is notoriously difficult, and is a requirement that has yet to be achieved on a space-based platform.

However, the tides may have begun to turn in interferometry's favour: long-baseline optical and infrared interferometers like *VLTI* and the *CHARA Array* have delivered transformative science in stellar physics, circumstellar environments, and exoplanet characterization[62,63], advanced by recent innovations in adaptive optics[64], fringe tracking[65], and integrated optics beam combiner[66], and radio arrays like *ALMA*, and up-coming next generation arrays like *SKA*, have begun to chip away the mystique that shrouds interferometry. Building on this momentum, the SPIFF Project would be the starting point for a long-term UK focus on space-based interferometry.

The path there starts with the plethora of existing science cases, many of which have already been developed from previous mission concepts, or the gap in available or planned facilities. Building on these, we propose to map these science goals to their associated enabling technologies, allowing us to identify technological blockers, areas requiring TRL raising, and opportunities to leverage other space-based use cases that align with UK priority areas[67], including Space Domain Awareness (SDA); In-orbit Servicing, Assembly, and Manufacturing (ISAM); and Positioning, Navigation, and Timing (PNT). We will also identify UK industrial partners whose technologies are mission for space-based interferometry, some of which have shown flight heritage for their respective sub-systems.

The output of this exercise would be the development of STMs and a technology trade study, where TRL-raising blockers and potential UK industrial partners have been identified. These results will feed into the next crucial step of SPIFF: a community workshop. Indeed, we believe it is ***paramount*** to bring the entirety of the UK astronomical community along for the ride. Therefore, we recommend hosting a community workshop where UK scientists and technologists can consolidate the results of the STMs and trade study to identify a common path forward.

This path would be via a survey of the opportunity landscape to build a technology demonstration mission. Such a mission would tightly focus on a few common, mission critical technologies, and prove space-based interferometry at the concept level, raising TRLs at a system level, and alleviate any lingering hesitation in the community. Indeed, a technology demonstrator would be ideally placed in the leadup to the US's Astro2030 Decadal Survey and ESA's next call for an M or L-class mission. Budget-friendly platforms like CubeSat and SmallSat would be suited for such a pathfinder mission, demonstrating mission critical technologies stabilising the optical bench, beam combination/fringe tracking, and other sub-systems that have not yet flown. showcasing UK-based capabilities.

The cornerstone of the SPIFF Project would be to leverage the expertise and lessons learned from previous mission concepts, working ***with*** the PIs and leaders from those efforts (***many of whom are co-signatories on this white paper***) to find a common path forward for space-based interferometry. We emphasize that we ***are not*** reinventing the wheel here. By establishing this international consortium of partners and collaborators, the UK puts itself front and centre as a leader in the field.

By the close of the Space Frontiers 2035 initiative, these research programs would culminate in momentous UK contributions to observatory-class proposals for a space-based interferometry mission, having provided flight heritage for mission critical systems and showing concept maturity. This will position the UK in a prime position to lead or provide significant technical contributions to such an observatory, should it be selected.

## 4. UK Leadership and Capability + Partnership Opportunities

*"…there is a limited amount of time for the UK to seize opportunities in space as other countries begin to 'outpace' us…"*

~ UK House of Lords, "The Space Economy: Act Now or Lose Out", 2025[67]

The UK is uniquely positioned to lead in space-based interferometry, with a rare trifecta of world-class expertise in ground-based interferometry and space instrumentation, a strong industrial base developing mission-critical technologies, and scientific leadership across the full range of relevant science cases. The UK previously led the *FISICA*[30,31,39] FP-7 Project, involving University College London, Cardiff University, RAL Space, and UK ATC—institutions with decades of heritage in infrared space instrumentation, including SPIRE for *Herschel*, MIRI for *JWST*, and the SPACEKIDS detector development program. UK universities also contribute significant ground-based interferometry experience through facilities such as *MROI* and *COAST*.

The UK industrial sector also brings essential technologies for future interferometer concepts. Oxford Space Systems have demonstrated flight heritage for deployable structures, including at scales required for space-based interferometry, which are an important technology and capability bridge towards larger science and commercial structures in orbit in the longer term[68]. The successful launch and formation-flying systems of *PROBA-3* have showcased UK capability in precision laser metrology system. Finally, far-IR optical components have been a specialty of the UK, including dating back to *Herschel*, and will be essential for any future far-IR mission. The SPIFF Project would map these capabilities, amongst others, to their associated science cases, identifying which common technologies are essential for a technology demonstrator mission.

The SPIFF project also benefits from strong international ties, as exemplified by our list of signatories, including signatories in Canada, Japan, and the United States. These countries have specifically been highlighted as priority partners by the UKSA, and SPIFF would strengthen these existing ties further, establishing the foundation for future multilateral consortium focused on long-term collaboration in space-based interferometry. Specifically, SPIFF aligns with national astronomy priorities recognized by the Canadian astronomy community and the CSA through recent long-range plan (LRP)[69] and on-going mid-term review (MTR). Furthermore, Japan has been identified as a key partner for further scientific collaboration in UK's recently published Modern Industrial Strategy[70]

By prioritizing space-based interferometry, the UK can position itself as a global partner of choice for future flagship missions, enabling leadership in science, mission architecture, and technology development. Such missions would sustain long-term employment across academia, public research institutions, and industry while maturing key enabling technologies within the UK. Decades of investment have created a critical window of opportunity; to seize it, the UK must now commit to space-based interferometry as a national priority.

## 5. Conclusion

*"Creating something new requires imagination directed at impossibility, not just the avoidance of impossibility."*
~ Westenberg, 2025

This white paper proposes a concept that should not, indeed ***cannot***, be considered in a vacuum. Many science cases across the EM spectrum would directly benefit from the high angular resolution observational capabilities unlocked by space-based interferometry, including those in other thematic areas of the Space Frontier 2035 initiative like heliophysics and solar system and planetary science. However, technical challenges still exist, and flight heritage has yet to be proven. Until now, the world has not yet been ready for space-based interferometry, but the day will soon come when it is. The limitations which have prevented the adoption of previous mission concepts should not cause further hesitation, whether they be technical challenges, concept immaturity, or perceived scepticism in the community. Rather, this should be a call to action. The SPIFF Project represents an opportunity for the UK to address these problems head on, learning from previous mission concepts, leveraging the UK's industrial base, and bringing the UK astronomical community along for the ride to alleviate lingering doubts in space-based interferometry. Culminating in a technology demonstration mission by the conclusion of the Space Frontiers 2035 program, the SPIFF Project will elevate the UK to being the partner of choice for future space-based interferometry missions, spanning the gamut from world-leading experts in science cases to domestic companies with mission critical technologies. The demands of observational data across the breadth of the astronomy and astrophysics community has made the capabilities of space-based interferometry a priority. There is an opportunity to be a world leader, if the UK **chooses** to step in, and we hope to have shown that with the UK's current expertise, it is entirely the UK's **choice** whether we want to sit on the sidelines while another country takes charge, or grasp the starring role for ourselves.

References

1. Kouveliotou, C. *et al.* Enduring Quests-Daring Visions (NASA Astrophysics in the Next Three Decades). Preprint at https://doi.org/10.48550/arXiv.1401.3741 (2014).
2. Unwin, S. C. & Beichman, C. A. Terrestrial Planet Finder: science overview. in *Optical, Infrared, and Millimeter Space Telescopes* vol. 5487 1216–1225 (SPIE, 2004).
3. Lawson, P. R., Lay, O. P., Johnston, K. J. & Beichman, C. A. *Terrestrial Planet Finder Interferometer Science Working Group Report*. *NASA STI/Recon Technical Report N* vol. 08 14326 https://ui.adsabs.harvard.edu/abs/2007STIN...0814326L (2007).
4. Cockell, C. S. *et al.* Darwin—A Mission to Detect and Search for Life on Extrasolar Planets. *Astrobiology* **9**, 1–22 (2009).
5. Lund, G. & Bonnet, H. DARWIN – The infrared space interferometer. *Comptes Rendus Académie Sci. - Ser. IV - Phys.* **2**, 137–148 (2001).
6. Quanz, S. P. *et al.* Large Interferometer For Exoplanets (LIFE): I. Improved exoplanet detection yield estimates for a large mid-infrared space-interferometer mission. *arXiv2101.07500* (2021).
7. Alei, E. *et al.* Large Interferometer For Exoplanets (LIFE): V. Diagnostic potential of a mid-infrared space-interferometer for studying Earth analogs. *Astron. Astrophys.* **665**, A106 (2022).
8. Torniamenti, S. *et al.* Optically thick winds of very massive stars suppress intermediate-mass black hole formation. Preprint at https://doi.org/10.48550/arXiv.2510.12465 (2025).
9. Giesers, B. *et al.* A detached stellar-mass black hole candidate in the globular cluster NGC 3201. *Mon. Not. R. Astron. Soc.* **475**, L15–L19 (2018).
10. Gaia Collaboration *et al.* Discovery of a dormant 33 solar-mass black hole in pre-release Gaia astrometry. *Astron. Astrophys.* **686**, L2 (2024).
11. Seeburger, R. *et al.* The physical properties of post-mass-transfer binaries. Preprint at https://doi.org/10.48550/arXiv.2511.13692 (2025).
12. Frost, A. J. *et al.* HR 6819 is a binary system with no black hole. Revisiting the source with infrared interferometry and optical integral field spectroscopy. *Astron. Astrophys.* **659**, L3 (2022).
13. Tacconi, L. *et al.* *Voyage 2050: Final Recommendations from the Voyage 2050 Senior Committee*. https://www.cosmos.esa.int/documents/1866264/1866292/Voyage2050-Senior-

Committee-report-public.pdf/e2b2631e-5348-5d2d-60c1-437225981b6b?t=1623427287109 (2021).

14. Gáspár, A. *et al.* Spatially resolved imaging of the inner Fomalhaut disk using JWST/MIRI. *Nat. Astron.* **7**, 790–798 (2023).

15. Williams, J. P. & Cieza, L. A. Protoplanetary Disks and Their Evolution. *Annu. Rev. Astron. Astrophys.* **49**, 67–117 (2011).

16. Wyatt, M. C. Evolution of debris disks. *Annu. Rev. Astron. Astrophys.* **46**, 339–383 (2008).

17. Kim, M., Kennedy, G. M. & Roccatagliata, V. The characterization of water ice in debris discs: implications for JWST scattered light observations. *Mon. Not. R. Astron. Soc.* **533**, 2801–2822 (2024).

18. Leisawitz, D. *et al.* The space infrared interferometric telescope (SPIRIT): High-resolution imaging and spectroscopy in the far-infrared. *Adv. Space Res.* **40**, 689–703 (2007).

19. Helmich, F. P. & Ivison, R. J. FIRI—A far-infrared interferometer. *Exp. Astron.* **23**, 245–276 (2009).

20. Linz, H. *et al.* Bringing high spatial resolution to the far-infrared. *Exp. Astron.* **51**, 661–697 (2021).

21. Leisawitz, D. *et al.* The Origins Space Telescope. in *UV/Optical/IR Space Telescopes and Instruments: Innovative Technologies and Concepts IX* vol. 11115 184–195 (SPIE, 2019).

22. Booth, M. *et al.* AtLAST Science Overview Report. Preprint at https://doi.org/10.48550/arXiv.2407.01413 (2024).

23. Wyatt, M. C. Resonant Trapping of Planetesimals by Planet Migration: Debris Disk Clumps and Vega's Similarity to the Solar System. *Astrophys. J.* **598**, 1321 (2003).

24. Bakx, T. J. L. C. *et al.* A warm ultraluminous infrared galaxy just 600 million years after the big bang. *Mon. Not. R. Astron. Soc.* **544**, 1502–1513 (2025).

25. Witstok, J. *et al.* Carbonaceous dust grains seen in the first billion years of cosmic time. *Nature* **621**, 267–270 (2023).

26. Ormerod, K. *et al.* Detection of the 2175 Å UV bump at z > 7: evidence for rapid dust evolution in a merging reionization-era galaxy. *Mon. Not. R. Astron. Soc.* **542**, 1136–1154 (2025).

27. Leisawitz, D. *et al.* The Space Interferometer for Cosmic Evolution (SPICE) Far-IR Probe. **55**, 160.04 (2023).

28. Spencer, L. D. *et al.* SPICE - The Space Interferometer for Cosmic Evolution: SPICE-ing up the Far-Infrared Universe. in *Optica Sensing Congress 2023 (AIS, FTS, HISE, Sensors, ES) (2023), paper FM4B.1* FM4B.1 (Optica Publishing Group, 2023). doi:10.1364/FTS.2023.FM4B.1.

29. Savini, G. & The FISICA-FP7 Consortium. Far-Infrared Space Interferometer Critical Assessment (FISICA): A double-Fourier space technology development. in *Imaging and Applied Optics* FM3D.4 (OSA, Arlington, Virginia, 2013). doi:10.1364/FTS.2013.FM3D.4.

30. Iafolla, V. A. *et al.* FISICA (Far Infrared Space Interferometer Critical Assessment) metrological problems and system requirements for interferometric observations from space. in *2014 IEEE Metrology for Aerospace (MetroAeroSpace)* 161–166 (2014). doi:10.1109/MetroAeroSpace.2014.6865913.

31. Savini, Giorgio & The FISICA-FP7 Consortium. Far Infra-red Space Intereferometer Critical Assessment: Scientific Definition and Technology Development for the Next Generation THz Space Interferometer | FP7. *CORDIS | European Commission* https://cordis.europa.eu/project/id/312818/reporting (2016).

32. Farrah, D. *et al.* Review: far-infrared instrumentation and technological development for the next decade. *J. Astron. Telesc. Instrum. Syst.* **5**, 020901 (2019).

33. Tielens, A. G. G. M. *The Physics and Chemistry of the Interstellar Medium*. (Cambridge University Press, 2005).

34. Sugahara, Y. *et al.* RIOJA. Complex Dusty Starbursts in a Major Merger B14-65666 at z = 7.15. *Astrophys. J.* **981**, 135 (2025).

35. Hashimoto, T. *et al.* Reionization and the ISM/Stellar Origins with JWST and ALMA (RIOJA): The Core of the Highest-redshift Galaxy Overdensity at z= 7.88 Confirmed by NIRSpec/JWST. *Astrophys. J. Lett.* **955**, L2 (2023).

36. Usui, M. *et al.* RIOJA. JWST and ALMA Unveil the Inhomogeneous and Complex Interstellar Medium Structure in a Star-forming Galaxy at z = 6.81. *Astrophys. J. Lett.* **991**, L38 (2025).

37. Stark, D. P., Topping, M. W., Endsley, R. & Tang, M. Observations of the first galaxies in the Era of JWST. in *Encyclopedia of Astrophysics (First Edition)* (ed. Mandel, I.) 453–499 (Elsevier, Oxford, 2026). doi:10.1016/B978-0-443-21439-4.00128-0.

38. Maiolino, R. *et al.* JADES: The diverse population of infant black holes at $4 < z < 11$: Merging, tiny, poor, but mighty. *A&A* **691**, (2024).

39. Savini, G. & The FISICA-FP7 Consortium. Far-Infrared Space Interferometer Critical Assessment (FISICA): A double-Fourier space technology development. in *Imaging and Applied Optics (2013), paper FM3D.4* (Optica Publishing Group, 2013). doi:10.1364/FTS.2013.FM3D.4.

40. Leisawitz, D. T. *et al.* SPECS: the kilometer-baseline far-IR interferometer in NASA's space science roadmap. in *Optical, Infrared, and Millimeter Space Telescopes* vol. 5487 1527–1537 (SPIE, 2004).

41. Carpenter, K. G., Schrijver, C. J. & Karovska, M. SI - The Stellar Imager. in (eds. Monnier, J. D., Schöller, M. & Danchi, W. C.) 626821 (Orlando, Florida , USA, 2006).

42. Rinehart, S. A. *et al.* The Space High Angular Resolution Probe for the Infrared (SHARP-IR). in *Space Telescopes and Instrumentation 2016: Optical, Infrared, and Millimeter Wave* vol. 9904 857–866 (SPIE, 2016).

43. de Graauw, T. & Team. FIR space heterodyne interferometer mission (ESPRIT). in *35th COSPAR scientific assembly* vol. 35 4564 (2004).

44. Rinehart, S. A. *et al.* A Long-Term Vision for Space-Based Interferometry. (2019).

45. Monnier, J. D. & endorsers, 67. A Realistic Roadmap to Formation Flying Space Interferometry. Preprint at https://doi.org/10.48550/arXiv.1907.09583 (2019).

46. Gendreau, K. C., Cash, W. C., Shipley, A. F. & White, N. E. MAXIM x-ray interferometry mission. in *Optics for EUV, X-Ray, and Gamma-Ray Astronomy* vol. 5168 420–434 (SPIE, 2004).

47. Duigou, J. M. L. *et al.* Pegase: a space-based nulling interferometer. in *Space Telescopes and Instrumentation I: Optical, Infrared, and Millimeter* vol. 6265 495–508 (SPIE, 2006).

48. Shao, M. *et al.* Space-based interferometric telescopes for the far infrared. in *Interferometry in Optical Astronomy* vol. 4006 772–781 (SPIE, 2000).

49. Danchi, W. C. & Lopez, B. The Fourier–Kelvin Stellar Interferometer (FKSI)—A practical infrared space interferometer on the path to the discovery and characterization of Earth-like planets around nearby stars. *Comptes Rendus Phys.* **8**, 396–407 (2007).

50. Linz, H. *et al.* InfraRed Astronomy Satellite Swarm Interferometry (IRASSI): Overview and study results. *Adv. Space Res.* **65**, 831–849 (2020).

51. Rinehart, S., Carpenter, K., van Belle, G. & Unwin, S. Interferometer evolution: imaging terras after building 'little' experiments (INEVITABLE). **9146**, 914617 (2014).

52. Leisawitz, D. & Rinehart, S. A. The path to far-IR interferometry in space: recent developments, plans, and prospects. in *Space Telescopes and Instrumentation 2012: Optical, Infrared, and Millimeter Wave* vol. 8442 749–761 (SPIE, 2012).

53. Monnier, J. D. *et al.* STARI: starlight acquisition and reflection toward interferometry. in *Space Telescopes and Instrumentation 2024: Optical, Infrared, and Millimeter Wave* vol. 13092 1132–1144 (SPIE, 2024).

54. Matsuo, T. *et al.* High spatial resolution spectral imaging method for space interferometers and its application to formation flying small satellites. *J. Astron. Telesc. Instrum. Syst.* **8**, 015001 (2022).

55. Ito, T. *et al.* SILVIA: Ultra-precision formation flying demonstration for space-based interferometry. *Publ. Astron. Soc. Jpn.* **77**, 1080–1089 (2025).

56. Rinehart, S. The balloon experimental twin telescope for infrared interferometry (BETTII). in vol. 7734 198–209 (SPIE, 2010).

57. Rinehart, S. *et al.* The balloon experimental twin telescope for infrared interferometry (BETTII): interferometry at the edge of the atmosphere. in *Optical and Infrared Interferometry IV* vol. 9146 914602 (SPIE, 2014).

58. Leisawitz, D. *et al.* The Japan-United States Infrared Interferometry Experiment (JUStIInE): balloon-borne pathfinder for a space-based far-IR interferometer. in vol. 12190 655–669 (SPIE, 2022).

59. National Academies of Sciences, Engineering, and Medicine. *Pathways to Discovery in Astronomy and Astrophysics for the 2020s*. (The National Academies Press, Washington, DC, 2023). doi:10.17226/26141.

60. Weiss, J. R., Smythe, W. D. & Lu, W. Science traceability. in 292–299 (IEEE, 2005).

61. Feldman, S. The science traceability matrix. in (2019).

62. Pedretti, E., Monnier, J. D., Ten Brummelaar, T. & Thureau, N. D. Imaging with the CHARA interferometer. *New Astron. Rev.* **53**, 353–362 (2009).

63. Eisenhauer, F., Monnier, J. D. & Pfuhl, O. Advances in optical/infrared interferometry. *Annu. Rev. Astron. Astrophys.* **61**, 237–285 (2023).

64. GRAVITY+ Collaboration. First Light for the GRAVITY+ Adaptive Optics: Extreme Adaptive Optics for the Very Large Telescope Interferometer. *ArXiv Prepr. ArXiv250921431* (2025).

65. Pedretti, E. *et al.* Robust determination of optical path difference: fringe tracking at the Infrared Optical Telescope Array interferometer. *Appl. Opt.* **44**, 5173–5179 (2005).

66. Monnier, J. D. *et al.* Imaging the surface of Altair. *Science* **317**, 342–345 (2007).

67. UK Engagement with Space Committee. *The Space Economy: Act Now or Lose Out*. https://publications.parliament.uk/pa/ld5901/ldselect/ldukspace/190/190.pdf (2025).

68. Fraux, Vincent, Reveles, Juan, & Lawton, Mike. AstroTube Max: The development and testing of a low mass extendible CFRP telescopic boom for satellite applications. in *Materials and Structures Symposium* (2015).

69. Gaensler, B. M. & Barmby, P. The Canadian Astronomy Long Range Plan 2020-2030. in vol. 235 290.01 (2020).

70. Department for Business and Trade. *The UK's Modern Industrial Strategy*. https://www.gov.uk/government/publications/industrial-strategy (2025).

List of Signatories

- Berke Vow Ricketti, Instrument Development Scientist, RAL Space, STFC
- Victoria Yankelevich, Instrument Development Scientist, RAL Space, STFC
- Emily Williams, Instrument Development Scientist, RAL Space, STFC
- Rebecca Harwin, Instrument Development Scientist, RAL Space, STFC
- David Pearson, Instrument Engineering Manager, RAL Space, STFC
- Ettore Pedretti, Calibration Scientist, RAL Space, STFC
- Andy Vick, Programme Lead Disruptive Technology, RAL Space, STFC
- Peter Huggard, Programme Lead Millimetre Wave Technology, RAL Space, STFC
- Chris Pearson, Programme Lead Astrophysics, RAL Space, STFC
- Gavin Dalton, Programme Lead Ground-based Astronomy RAL Space, STFC
- Will Grainger, Systems Engineering Group Lead, RAL Space, STFC
- Damien Weidmann, Spectroscopy Group Lead, RAL Space, STFC
- Alistair Glasse, Senior Instrument Scientist, ROE, UK ATC, STFC
- Olivia Jones, STFC Webb Fellow, ROE, UK ATC, STFC
- Christophe Dumas, Director of UK ATC, ROE, UK ATC, STFC
- Charles Cockell FRSE, Professor of Astrobiology, U. Edinburgh
- Sebastian Marino, Associate Professor of Astrophysics, U. Exeter
- Andrew Blain, Prof. of Observational Astronomy, U. Leicester
- Giorgio Savini, Prof. of Astrophysics, UCL
- Ziri Younsi, Lecturer in Astrophysics, UCL
- Renske Smit, STFC Ernest Rutherford Fellow, LJMU
- Sebastian Kamann, UKRI Future Leaders Fellow, LJMU
- Tim D. Pearce, Stephen Hawking Fellow, U. Warwick
- Chris Benson, Postdoctoral Research Associate, Cardiff Hub for Astrophysics Research and Technology, Cardiff U.
- Matthew Smith, Director of Postgraduate Research Studies, Cardiff Hub for Astrophysics Research and Technology, Cardiff U.
- Stephen Eales, Co-Director of the Cardiff Hub for Astrophysics Research and Technology, Cardiff U.
- Boon-Kok Tan, Senior Researcher, U. Oxford
- Suzanne Aigrain, Professor of Astrophysics & Fellow of All Souls College, U. Oxford
- Jayne Birkby, Professor of Astrophysics & Fellow of Brasenose College, U. Oxford
- Mark Wyatt, Co-Director of the Institute for Astronomy, U. Cambridge
- Jan Forbrich, Associate Professor, Centre for Astrophysics Research, U. Hertfordshire / External Research Associate, Centre for Astrophysics, Harvard & Smithsonian
- Chris Bee, Business Development Director, Oxford Space Systems
- Taro Matsuo, Professor of Astrobiology, U. Osaka, Japan
- Hiroshi Matsuo, Associate Professor, NAOJ, Japan
- Locke Spencer, Dept. Chair and Associate Professor of Physics & Astronomy, U. Lethbridge, Canada
- David Naylor, Professor and Board of Governors' Chair Emeritus, U. Lethbridge / Blue Sky Spectroscopy Inc., Founder and CTO
- Gerard van Belle, Ad Astra Space LLC, USA
- Lee Mundy, Professor of Astronomy, U. Maryland, USA
- David T. Leisawitz, Associate Chief of the Exoplanets and Stellar Astrophysics Laboratory, NASA Goddard Space Flight Center, USA